\documentclass{aa}  

\usepackage{graphicx}
\usepackage{txfonts}

\usepackage{epstopdf}

\usepackage{epsfig}
\usepackage{url}

\usepackage{upgreek}  
\newcommand\unit[1]{\,{\rm #1}} 

\def\lta{~\raise.4ex\hbox{$<$}\llap{\lower.6ex\hbox{$\sim$}}~}
\def\gta{~\raise.4ex\hbox{$>$}\llap{\lower.6ex\hbox{$\sim$}}~}
\def\eg{{\it e.g.}}
\def\ie{{\it i.e.}}

\def\msol{\ensuremath{\rm \mass_\odot}} 
\newcommand\solm\msol

\begin{document}

\title{Comet 67P/Churyumov-Gerasimenko: Constraints on its origin from OSIRIS observations}
\titlerunning{Constraints on 67P origin}

\author{H. Rickman\inst{1,2}\and S. Marchi\inst{3} \and M. F. A'Hearn\inst{4} C. Barbieri\inst{5} \and M. R. El-Maarry\inst{6} \and  C. G\"uttler\inst{7} \and W.-H. Ip\inst{8} \and H. U. Keller\inst{9} \and P. Lamy\inst{10} \and F. Marzari\inst{5} \and M. Massironi\inst{11,12} \and G. Naletto\inst{12, 13, 14} \and M. Pajola\inst{12} \and H. Sierks\inst{7} \and D. Koschny\inst{15} \and R. Rodrigo\inst{16,17} \and M. A. Barucci\inst{18} \and J.-L. Bertaux\inst{19} \and I. Bertini\inst{12} \and G. Cremonese\inst{20} \and V. Da Deppo\inst{14} \and S. Debei\inst{21} \and M. De Cecco\inst{22} \and S. Fornasier\inst{18} \and M. Fulle\inst{23} \and O. Groussin\inst{10} \and P. J. Guti\'errez\inst{24} \and S. F. Hviid\inst{25} \and L. Jorda\inst{10} \and J. Knollenberg\inst{25} \and J.-R. Kramm\inst{7} \and E. K\"uhrt\inst{25} \and M. K\"uppers\inst{26} \and L. M. Lara\inst{24} \and M. Lazzarin\inst{5} \and J. J. Lopez Moreno\inst{24} \and H. Michalik\inst{27} \and L. Sabau\inst{28} \and N. Thomas\inst{6} \and J.-B. Vincent\inst{7} \and K.-P. Wenzel\inst{15} }


\authorrunning{Rickman et al.}

\institute{P.A.S. Space Research Center, Bartycka 18A, PL-00-716 Warszawa, Poland
\and Dept. of Physics and Astronomy, Uppsala University, Box 516, SE-75120 Uppsala, Sweden
\and Southwest Research Institute, 1050 Walnut St., Boulder, CO 80302, USA
\and Department for Astronomy, University of Maryland, College Park, MD 20742-2421, USA
\and Department of Physics and Astronomy "G. Galilei", University of Padova, Vic. Osservatorio 3, 35122 Padova, Italy
\and Physikalisches Institut, Sidlerstrasse 5, University of Bern, CH-3012 Bern, Switzerland
\and Max-Planck Institut f\"ur Sonnensystemforschung, Justus-von-Liebig-Weg 3, 37077 G\"ottingen, Germany
\and Institute for Space Science, National Central University, 32054 Chung-Li, Taiwan
\and Institute for Geophysics and Extraterrestrial Physics, TU Braunschweig, 38106 Braunschweig, Germany
\and Aix Marseille Universit\'e, CNRS, LAM (Laboratoire d'Astrophysique de Marseille) UMR 7326, 13388 Marseille, France
\and Dipartimento di Geoscienze, University of Padova, via G. Gradenigo 6, 35131 Padova, Italy
\and Centro di Ateneo di Studi ed Attivit\'a Spaziali, "Giuseppe Colombo" (CISAS), University of Padova, via Venezia 15, 35131 Padova, Italy
\and Department of Information Engineering, University of Padova, via G. Gradenigo 6/B, 35131 Padova, Italy
\and CNR-IFN UOS Padova LUXOR, via Trasea 7, 35131 Padova, Italy
\and Scientific Support Office, European Space Agency, 2201 Noordwijk, The Netherlands
\and Centro de Astrobiologia (INTA-CSIC), 28691 Villanueva de la Canada, Madrid, Spain
\and International Space Science Institute, Hallerstrasse 6, CH-3012 Bern, Switzerland
\and LESIA, Obs. de Paris, CNRS, Univ Paris 06, Univ. Paris-Diderot, 5 pl. J. Janssen, 92195 Meudon, France
\and LATMOS, CNRS/UVSQ/IPSL, 11 bd d'Alembert, 78280 Guyancourt, France
\and INAF - Osservatorio Astronomico, vicolo dell'Osservatorio 2,  35122 Padova, Italy
\and Department of Mechanical Engineering, University of Padova, via Venezia 1, 35131 Padova, Italy
\and UNITN, Univ. di Trento, via Mesiano 77, 38100 Trento, Italy
\and INAF - Osservatorio Astronomico, via Tiepolo 11, 34143 Trieste, Italy
\and Instituto de Astrofisica de Andaluc\'{\i}a (CSIC), Glorieta de la Astronom\'{\i}a s/n, 18008 Granada, Spain
\and Institute of Planetary Research, DLR, Rutherfordstrasse 2, 12489 Berlin, Germany
\and Operations Department, European Space Astronomy Centre/ESA, P.O. Box 78, 28691 Villanueva de la Canada, Madrid, Spain
\and Institut f\"ur Datentechnik und Kommunikationsnetze der TU Braunschweig, Hans-Sommer-Str. 66, 38106 Braunschweig, Germany
\and Instituto Nacional de T\'ecnica Aeroespacial, Carretera de Ajalvir, p.k. 4, 28850 Torrejon de Ardoz, Madrid, Spain
 }

\abstract{One of the main aims of the ESA Rosetta mission is to
  study the origin of the solar system by exploring comet
  67P/Churyumov-Gerasimenko at close range.} 
 {In this paper we discuss the origin and evolution of comet
  67P/Churyumov-Gerasimenko in relation to that of comets
  in general and in the framework of current solar system formation models.} 
 {We use data from the OSIRIS scientific cameras as basic
   constraints. In particular, we discuss the overall bi-lobate shape
   and the presence of key geological features, such as layers and
   fractures. We also treat the problem of collisional evolution of
   comet nuclei by a particle-in-a-box calculation for an estimate of
   the probability of survival for 67P/Churyumov-Gerasimenko during
   the early epochs of the solar system.} 
{We argue that the two lobes of the 67P/Churyumov-Gerasimenko nucleus
  are derived from two distinct objects that have formed a contact
  binary via a gentle merger. The lobes are separate bodies, though sufficiently similar
  to have formed in the same environment. An estimate of the
  collisional rate in the primordial, trans-planetary disk shows that
  most comets of similar size to 67P/Churyumov-Gerasimenko are likely
  collisional fragments, although survival of primordial planetesimals 
  cannot be excluded.} 
{A collisional origin of the contact binary is suggested, and the low
  bulk density of the aggregate and abundance of volatile species show
  that a very gentle merger must have occurred. We thus consider
  two main scenarios: the primordial accretion of planetesimals,
  and the re-accretion of fragments after an energetic impact onto a
  larger parent body. We point to the primordial
  signatures exhibited by 67P/Churyumov-Gerasimenko and 
  other comet nuclei as critical tests of the collisional evolution.}  


\date{[Received / accepted]}
 
\maketitle

\section{Introduction}\label{sec:intro}

Promoting the quest for the origin of the solar system is fundamental
to the concept of the ESA/Rosetta mission. A comet nucleus --
specifically, that of 67P/Churyumov-Gerasimenko (hereinafter 67P) --
was chosen as the target because of the common notion that comet nuclei
are icy planetesimals formed beyond the snow line in the nascent solar
system and the evidence that their atomic and molecular compositions
provide a good match to the primordial material out of which the solar
system was built. In particular, comet 67P was selected as the back-up
target after the failure to launch Rosetta toward 46P/Wirtanen, partly
because it was expected to be relatively fresh after having
recently encountered Jupiter (in 1959) and thus had its perihelion
distance much reduced.

The formation of the icy planetesimals is currently viewed within
either of two different scenarios. One is the classical model of
hierarchical accretion (Weidenschilling \cite{wei08}), and the other
considers the assembly of gravitationally unstable ``pebble clouds''
by gas-grain instabilities in the solar nebula (Johansen et
al. \cite{joh14}). Recent simulations (Wahlberg Jansson and Johansen
\cite{wah14}) have shown how planetesimals in the size range from 10
to 1000\unit{km} can thus be produced.

Independent of which scenario is relevant to comet 67P, it seems clear
that its ultimate origin has to be sought in the same, very early
environment. The exact place is more difficult to specify. The region
inside the initial orbits of the giant planets can likely be
discarded, since cometary ices would not condense there. Because of
the gas drag presented by the solar nebula, km-sized planetesimals
formed in the giant planet zone are unlikely to be gravitationally
scattered outward in order to populate the well-known cometary
reservoirs (Brasser et al. \cite{bra07}). Thus, the trans-planetary
space, hosting a massive disk of planetesimals, is currently favored
as the birth place of comet nuclei.

The issue of collisional evolution among those planetesimals has been
discussed for almost 20 years.  Davis and Farinella (\cite{dav97})
modeled the collisional evolution of the Edgeworth-Kuiper Belt (EKB)
and argued that its members with radii of a few km -- like typical
Jupiter family comets -- are multigenerational fragments formed by the
splitting of larger objects. Stern and Weissman (\cite{ste01})
considered the formation of the Oort cloud out of planetesimals formed
in the giant planet zone. They showed that, during the course of this
gravitational scattering process, comet nuclei would be destroyed by
collisions with small debris. However, following the introduction of
the Nice Model for the evolution of the solar system (Tsiganis et
al. \cite{tsi05}), the EKB is no longer thought to be the direct
survivor of the initial comet population\footnote{The cold EKB may
  indeed be a remnant of the primordial disk, but it carries a very
  small fraction of the total mass (Fraser et al. \cite{fra14}).}, and
according to Brasser and Morbidelli (\cite{bra13}), the actual Oort
cloud formation may have little to do with the classical concept of
scattering of planetesimals from the formation zone of the giant
planets.

In this paper we discuss some of the evidence that OSIRIS imaging of
the 67P nucleus may bring regarding its origin and evolution. We argue
that the bi-lobate shape is likely due to a gentle merger of the two
lobes as separate bodies in a low-velocity collision. This conclusion 
is supported by additional observations, such as the
presence of thick layering. We place this into the framework of the
collisional evolution issue as it currently stands, and we highlight
the importance of resolving the remaining issues about the fundamental
question, whether the 67P nucleus is the result of an early merging of
two distinct planetesimals or was re-accreted in connection with a
major collision long after the planetesimals were formed. Meanwhile,
we also point out that, in any case, this nucleus does not appear to
have completely lost its primordial signatures and may thus retain its
significance as a witness of planetesimal formation in the nascent
solar system.

In Sect.~\ref{sec:shape} we discuss the role of past erosion in shaping 
the 67P nucleus and evidence regarding its bi-lobate shape and the 
interpretation as a contact binary. We summarize the evidence 
concerning structural features of the nucleus and the origin of the 
contact binary in a low-velocity collision. Sect.~\ref{sec:colli} gives an
estimate of the collision rate in the trans-planetary disk, which
illustrates that it is generally unlikely for km-sized planetesimals
to survive intact during the expected lifetime of this disk, and in
Sect.~\ref{sec:escape} we discuss possible ways for comets to escape
destruction in the event that 67P should prove to be such a survivor.
Our conclusions are summarized in Sect.~\ref{sec:conclu}.

\section{Overall shape and structural features of 67P}\label{sec:shape}

\subsection{Previous erosion}

\begin{figure*}[h]
\includegraphics[width=12cm]{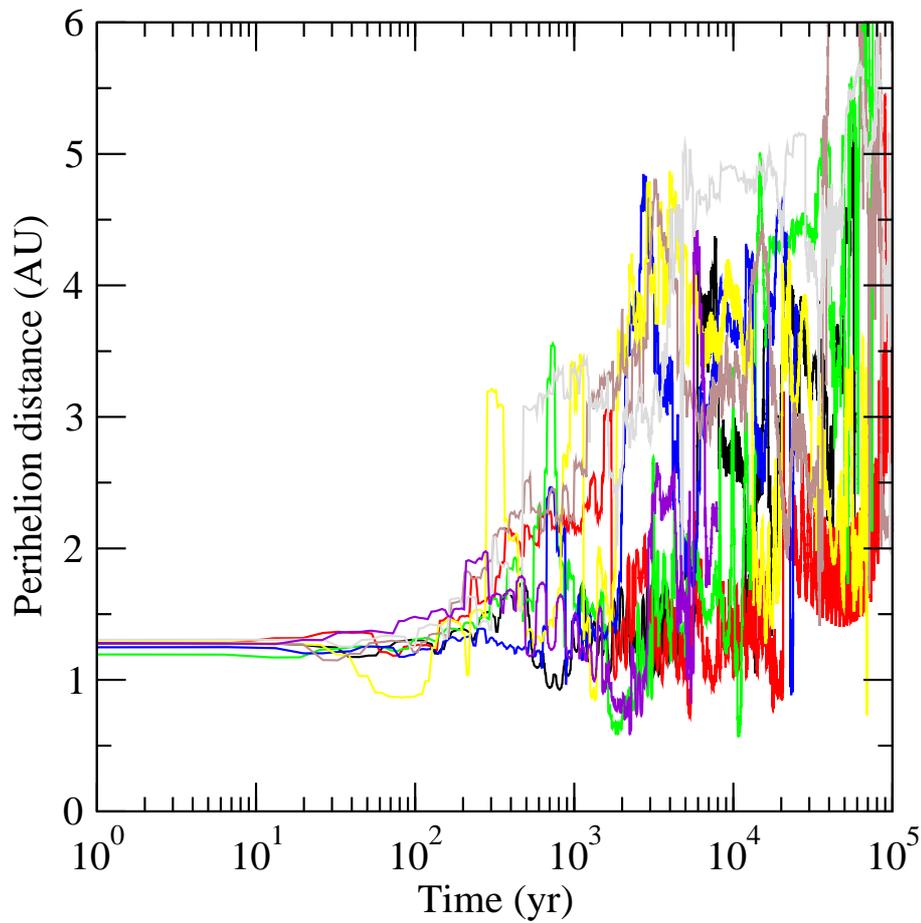}
\caption{Evolutions of the perihelion distance of 67P, found by forward integrations 
of random initial orbits leading into current orbits close to the one of 67P. The time 
plotted on the abscissa is counted backward from the present. The different colors 
mark the different examples chosen. We note that these are not clones of 67P, and the 
recent history of 67P (including the close encounter with Jupiter in 1959) is not 
reproduced. The focus is on the long-term dynamics, shown in the right part of the 
diagram. Chaotic divergence of the backward motions sets in, mostly a few centuries 
ago, due to close encounters with Jupiter. Courtesy Ryszard Gabryszewski.}
\label{fig_rysz}
\end{figure*}

Thermo-physical models reveal that 67P could have lost a surface layer
of up to several hundred meters thickness due to the accumulated
activity (Sierks et al. \cite{sie15}; Keller et al. \cite{kel15})
during its previous orbits in the inner solar system. This estimate is
based on the median number of such orbits as estimated from long-term
integrations for fictitious objects similar to 67P, and the specific
evolutions of the perihelion distance of the comet, which largely
governs the erosion rate. Figure~\ref{fig_rysz} shows an example of
the large variety of previous evolutions of the perihelion distance of
67P that may lead into orbits close to the present one according to
long-term integrations. In general, these evolutions have brought the
comet's perihelion closer to the Sun, but Figure~\ref{fig_rysz} covers 
only part of the whole time spent by the comet in the inner solar system. 
In many cases, the comet had made earlier visits into orbits with small 
perihelion distance, which also contributed to the cumulative erosion. 
We note that the dynamical time scale found by these integrations may be 
consistent with the capture time scale recently estimated by Guzzo and 
Lega (\cite{gul15}), although further work is needed to confirm this.

While the use of the current shape and spin axis orientation cannot
have a drastic influence on those results, the extrapolation of the
current activity level may possibly be an important source of
error. In particular, there is a concern about the possibility that
the comet has undergone long periods of dormancy (Kres\'ak
\cite{kre87}), when its gas production was close to zero.  This has
been suggested for individual comets -- for instance, Levison et
al. (\cite{lev06}) found that the orbital properties of 2P/Encke may
be explained, if this comet spent a long time in the past as a dormant
comet and was relatively recently woken up, as its perihelion distance
was decreased by a secular resonance.

In case dormancy can become permanent, \ie, comets stay dormant until 
they are dynamically ejected, this may 
influence our estimate of past erosion for the reason that the current
activity of 67P would indicate the comet to be relatively young in the
dynamical sense. This youth is of course not verified by long-term
orbital integrations due to the strongly chaotic behavior, \ie, short
Lyapunov time scales, of Jupiter family comets.  Permanent dormancy
would seem to be supported by the work of Duncan and Levison
(\cite{dun97}), who concluded that the physical lifetime of these
comets must be limited because their orbits retain a memory of the
very low inclinations in the scattered disk that they assumed as the
source for captures. However, the scattered disk is now recognized to
be much more excited, so this argument may not be of relevance any
more.

We tentatively conclude that a thick surface layer has likely been
eroded from the 67P nucleus by previous gas-producing activity,
although the above-mentioned estimate of several hundred meters should perhaps be
regarded as an upper limit. In any case, this cautions that
activity-driven erosion may result in complex landscapes (Malin and
Zimbelman \cite{mal86}; Thomas et al. \cite{tho05}), which could
affect our ability to interpret geomorphological features in terms of
primordial processes. This is particularly true for equatorial regions
and the southern hemisphere due to their computed higher erosion rates
(Keller et al. \cite{kel15}). Furthermore, it should be noted that
the small-scale details seen on the surface of the 67P nucleus are
likely of very recent origin in most cases. For instance, because
67P currently crosses the main asteroid belt, following the methodology
developed by Marchi et al. (\cite{mar10}, \cite{mar12}a,b) we estimate that
some $100-1000$ craters larger than a meter should have
formed over a time scale of 10$^3$\unit{y} -- however, these craters are not observed.

For these reasons, we will focus our attention on the overall
bi-lobate shape and the large-scale structural properties of two key
regions, {\it Seth} and {\it Hathor} (Thomas et al. \cite{tho15}),
which are least affected by activity-driven erosion (Keller et
al. 2015). These regions face the ``neck'' from the larger and 
smaller lobes, respectively.

\subsection{Contact binary}

The bi-lobate shape of 67P has important implications for the
formation of the nucleus. Computations based on the current shape,
spin axis orientation, and a homogeneous composition show that the
neck-facing regions (in particular, the Seth and Hathor regions) are
not subject to an increased dose of solar radiation that would imply
preferential erosion (Sierks et al. \cite{sie15}; Keller et
al. \cite{kel15}). Although it cannot be ruled out that 67P had a
radically different spin axis orientation in the past, it seems
unlikely that erosion of a homogeneous object alone could explain the
current shape of 67P. It is nevertheless possible that a very
heterogeneous nucleus (with locally higher concentration of volatiles,
or different terrain properties) could result in preferential, local
erosion. Such a heterogeneous structure, however, may require mixing
of planetesimals with different volatile compositions or abundances,
probably only attainable through collision and merging of objects
formed at different heliocentric distances.

Alternatively, the bi-lobate shape may strongly suggest 67P to be a
contact binary. While contact binaries have been advocated for
elongated comet nuclei (\eg, 1P/Halley, 19P/Borrelly and 103P/Hartley
2), the only unquestionable case remains that of 8P/Tuttle based on
radar imaging (Harmon at al. \cite{har10}) and indirectly confirmed by
HST observations (Lamy et al. \cite{lam08}).  Indeed, based on its
appearance, 67P is a highly likely candidate. As such, however, its
shape appears unusual, since the two lobes have their longest axes
nearly perpendicular whereas they are aligned in most if not all other
contact binaries.

Binary systems are ubiquitous in various small-body populations,
including the near-Earth asteroids (NEAs), main-belt asteroids,
Jupiter Trojans, Centaurs, and Trans-Neptunian Objects (TNOs). Contact
binaries alone are estimated to constitute $10-20$\% of all NEAs,
Trojans, and TNOs (Sheppard and Jewitt \cite{she04}; Noll et
al. \cite{nol08}; Mann et al. \cite{man07}; Benner et
al. \cite{ben06}, \cite{ben08}). Several mechanisms for the formation
of binary systems have been proposed, ranging from fission of small
rapidly rotating NEAs (Scheeres \cite{sch07}), to the formation of
large binary TNOs by gravitational collapse (Nesvorn\'y et
al. \cite{nes10}). Whether or not these processes are applicable to
highly porous, small comets deserves further investigation.

\subsection{Structural features}

Concerning possible structural features, the Seth region displays
multiple semi-planar terrains, or facets (Thomas et al.
\cite{tho15}; Massironi et al. \cite{mas15}). Planar terrains are, in some cases, reminiscent of what
has been seen on other comet nuclei (\eg, Belton et
al. \cite{bel07}). A key difference, however, is that Seth is
characterized by the presence of deep collapsed pits (Vincent et
al. \cite{vin15}), whose dust-free walls allow us to glance at the
local vertical stratigraphy. In many instances, these walls reveal
quasi-parallel lineaments and small terraces, which may be interpreted
as the surface expression of internal layering (Marchi et
al. \cite{mar15}; Massironi et al. \cite{mas15}). 

\begin{figure*}[h]
\includegraphics[width=15cm]{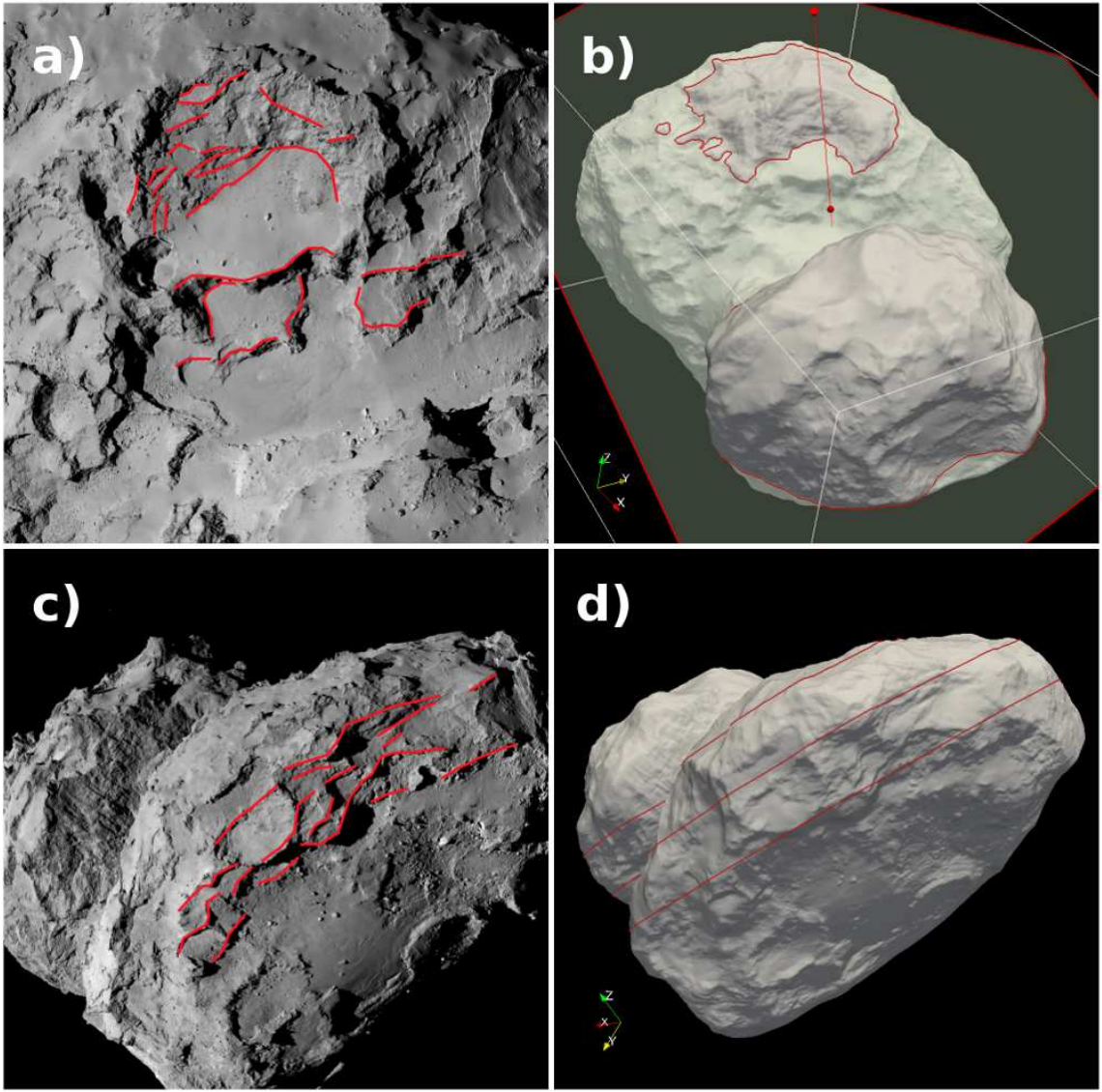}
\caption{Large-scale layering on the large lobe. a) Close-in view of
  the pitted terrain on the large lobe (Seth region). The red lines
  indicate some of the prominent lineaments, interpreted as surface
  expression of layering. b) A plane indicating the prevailing local
  orientation of the layers for the largest pit in the Seth region. c)
  A set of lineaments and high topographic reliefs on the opposite
  side of the large lobe. d) Same plane as in b) and two additional
  parallel planes. The orientation of these planes matches the
  lineaments and topographic reliefs, suggesting the latter are
  surface expressions of internal layering. We note that the local
  orientation of the plane may vary from place to place. For
  simplicity we assumed that the best fit plane does not change
  attitude within the large lobe. However, a similar conclusion
  applies for bending following the local convex shape, or along the
  y-axis in panel b.}
\label{fig_seth}
\end{figure*}

In the case of the largest terrace in the Seth region, a shape model
(Jorda et al. \cite{jor15}) was used to fit the floor of the pit with
a plane to obtain an approximate local orientation of the putative
layer (Fig.~\ref{fig_seth}b).  Interestingly, the orientation of the
resulting plane, extrapolated to the extreme end of the large lobe,
matches the orientation of a topographic ridge used to mark the
boundary of two very morphologically distinct terrains (Ash and
Imhotep regions; Thomas et al. \cite{tho15}), suggesting that the flat
floors and lineaments are truly indicative of internal layering, with
an estimated local depth of at least a few hundred meters (see
Marchi et al. \cite{mar15}; Massironi et al. \cite{mas15}).

\begin{figure*}[h]
\includegraphics[width=15cm]{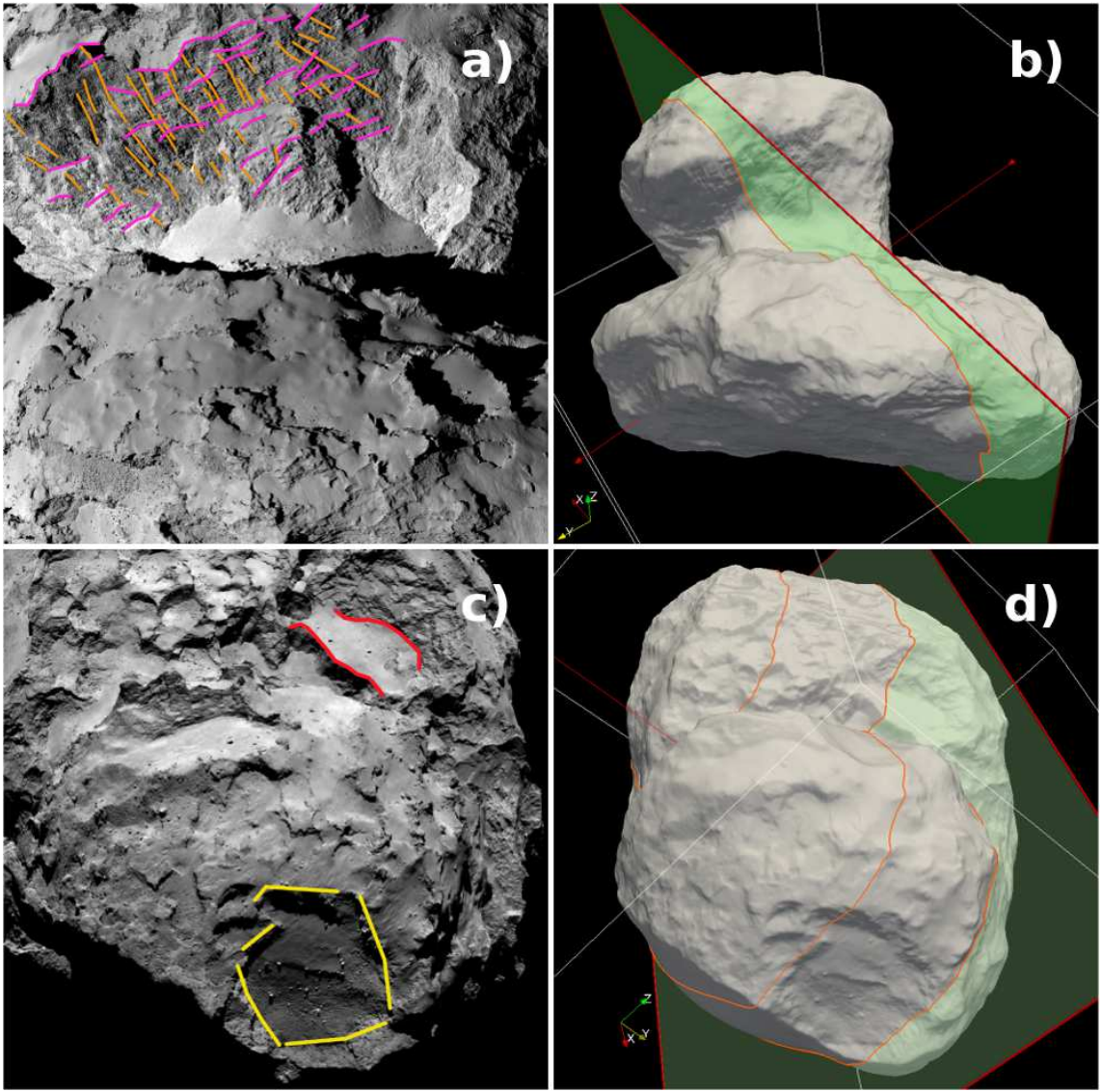}
\caption{Large-scale fractures on the small lobe. a) Cross-cutting
  pattern of fractures (orange lines) and layers (purple lines) on the
  Hathor region. b) A plane indicating the prevailing orientation of
  the fractures overlaid on a shape model. c) A view of the top of the
  small lobe, with some major features highlighted: Layering on the
  large lobe (red lines; see Fig.~\ref{fig_seth}), rim of the Hatmehit
  depression (yellow lines). d) The same plane shown in b), and an
  additional parallel plane. The orientation of these planes matches
  the direction of two sides of the Hatmehit depression.}
\label{fig_hathor}
\end{figure*}

The Seth region faces a prominent dust-free cliff in the small lobe,
named Hathor, which reveals the ubiquitous presence of fractures and
terraces, the latter being interpreted as the result of layering
(Fig.~\ref{fig_hathor}a). The two sets of layering in the Seth and
Hathor regions have local spatial orientations that are not compatible
with being the expression of a homogeneous layering wrapping the two
lobes (Marchi et al. \cite{mar15}; Massironi et al. \cite{mas15}). A
similar conclusion is also reached by looking at the global
distribution of semi-planar terrains on the surface imaged so far
(Massironi et al.  \cite{mas15}).

The Hathor region also hosts another large-scale geological feature
with possible implications for the formation of 67P, \ie, an
impressive set of fractures that run for hundreds of meters from the
base of the small lobe to its summit (Fig.~\ref{fig_hathor}a). The
largest of these are inferred to be several meters wide, based on the
cast shadows. Several hypotheses for their formation have been
investigated, such as thermal fracturing, and gradual sand-blasting
erosion due to activity from the neck region. Thermal fracturing
typically results in polygonal patterns, such as those observed in
permafrost on Mars and Earth (\eg, Levy et al. \cite{lev10}), and is
therefore an unlikely explanation for this well-organized set of large
fractures.  Sand-blasting from a cometary jet generally exerts a
pressure much lower than the relevant material strength, and therefore
also does not appear to be a likely mechanism.

Alternatively, wide spread fracturing could be achieved through
large-scale torques due to re-orientation of the two lobes, or by
rotational stresses exceeding the failure strength. However, such
torque-induced fractures are expected to form in the neck region
itself, where most of the resulting stresses would concentrate, rather
than the neck-facing regions. Moreover, given the almost symmetric
nature of the Seth and Hathor regions with respect to the neck, such
processes should result in similar fracturing. While some degree of
fracturing is indeed observed in the Seth region, the lack of an
extensive and well-organized system of fractures suggests that the two
lobes may have distinct properties.

In addition, a spatial analysis of the orientations of the fractures
suggests they penetrate deeply into the small lobe, indicating that they 
may be pervasive. Support to this
conclusion comes from the morphology of a large depression, $\sim$1\unit{km}
across, at the summit of the small lobe, named Hatmehit region (Thomas
et al. \cite{tho15}). Whether this depression is the result of a
collapse or an outburst, its rhombic shape indicates the presence of
planes of weakness within the small lobe, which appear to be aligned
with the Hathor fracture system (Fig.~\ref{fig_hathor}c-d).  It
therefore seems possible that these fractures are structural features,
pervading the smaller lobe.

\subsection{Binary formation}

The above geomorphological evidence on 67P supports the view that the
bi-lobate nature of 67P is due to collision and merging of two
distinct objects. Further constraints on this event can be derived
from the high porosity of 67P (Sierks et al. \cite{sie15}) and the
content of volatile species. The high porosity requires peak pressures
likely not exceeding $\sim100$\unit{kPa} (Yasui and Arakawa \cite{yas09}), and
hence an accretional speed not exceeding a few tens of m/s. The
average impact speed of small planetesimals is thought to have evolved
from $\sim10$\unit{m/s} during the early stages still dominated by gas
(Kokubo and Ida \cite{kok00}) to $\sim500$\unit{m/s} corresponding to
viscous stirring of the planetesimal disk by its largest members
(Levison et al. \cite{lev11}). These considerations narrow down two
possible formation scenarios for 67P.

In one case, 67P is the result of a low-speed collision between two
planetesimals, which most likely took place in the earliest stages of
the disk evolution before gas dispersal. This would mean that the
structure of 67P dates back to the time of planetesimal formation and
has not been significantly altered by later evolution. Alternatively,
67P is the result of re-accumulation of fragments ejected during a
catastrophic collision involving a larger planetesimal. Numerical
simulations (\eg, Jutzi et al. \cite{jut10}) have shown that the
catastrophic disruption of porous parent bodies may result in ejection
velocities $\lta10$\unit{m/s} (comparable to the mutual impact speed
among fragments), while for a typical impact speed of
$\sim500$\unit{m/s} or lower a significant volume of the parent body
may experience low shock pressures (\eg, Leinhardt and Stewart
\cite{lei09}, rescaled to lower impact speeds). In addition, a highly
porous parent body will greatly help dumping the shock wave,
therefore reducing the pressure exerted far from the impact location
(W{\"u}ennemann et al. \cite{wue06}). Therefore, the collisional scenario
should also be kept in consideration as a possible mode of origin of
67P. In this case too, the structure formed during the re-accumulation
must not have been destroyed by subsequent evolution, but preceding
collisions on the same parent body cannot be excluded.

Our discussion suggests that the two lobes of 67P are two distinct
objects that merged in a low-speed collision. However, the issue
remains, whether this collision was part of the primordial
planetesimal accretion, or if it involved fragments produced later
by a collision experienced by a larger parent body.  This is very
important, not only to realize which scenario would best explain the
properties of 67P as observed by OSIRIS and other instruments, but
also because current concepts on solar system evolution have important
implications for the collisional evolution of comets. We shall now
turn to the latter.

\section{Collisional evolution in the trans-planetary disk}\label{sec:colli}

\subsection{General framework}

We place our discussion within the current paradigm of solar system
evolution as defined by the Nice Model (Tsiganis et al. \cite{tsi05};
for the latest development see Levison et al. \cite{lev11}). This
model features a disk of planetesimals beyond the initial, planetary
orbits, which should have extended roughly from 15 to 30\unit{AU} in
heliocentric distance. The disk was formed in the nascent solar
system, when the planetesimals were formed, and it was dispersed as a
consequence of planet migration following the dynamical instability of
the giant planets.

The dispersal is thought to have taken place about $4.1-4.2$\unit{Gy}
ago (Morbidelli et al. \cite{mor12}; Marchi et al. \cite{mar13}),
implying $\sim 0.4$\unit{Gy} of prior collisional evolution. The total
mass of the disk is constrained to be in the range $20-50$ Earth
masses (Tsiganis et al. \cite{tsi05}). The timing of the dynamical
instability is assumed to coincide with the start of the Late Heavy
Bombardment (LHB).  Among the important effects of the instability,
there is also the gravitational scattering of objects into the Oort
cloud and the scattered disk (Brasser and Morbidelli \cite{bra13}). We
consider that, at the typical velocities of collisions in the
trans-planetary disk, projectiles with diameter ($D$) larger than 1 km
would shatter targets of the same size as 67P ($D \sim 4$\unit{km},
assuming a past erosion of a $300$\unit{m} thick layer). This estimate
is likely conservative, judging from the expected material properties
(\eg, Benz and Asphaug \cite{ben99})\footnote{Morbidelli and Rickman
  (\cite{mor15}) used the Benz and Asphaug (\cite{ben99}) scaling law
  for hard ice, hit at 1~\unit{km/s}, and found that the minimum
  diameter of a disruptive projectile for a target of the same size as
  67P was always about 1~\unit{km} or less throughout the primordial
  disk. However, for porous targets the minimum projectile diameter
  would be smaller, using the weak ice scaling law of Leinhardt and
  Stewart (\cite{lei09}).}.

Concerning the availability of such projectiles, a recent study by
Johansen et al. (\cite{joh15}) of planetesimal formation in a pebble
cloud found the cumulative size frequency distribution (SFD) to be
quite shallow with a power law index $\alpha \simeq -1.8$ for $ 20
\lesssim D \lesssim 200$\unit{km}, although the applicability of this
shallow slope down to km-sized objects needs to be further
investigated. Additional constraints on this size range come from the
observed trans-Neptunian populations and Trojan asteroids (the current
leftovers of the primordial disk).  These populations exhibit
two-sloped cumulative SFDs, with a break at $D \sim 100$\unit{km}.
The slope of the cumulative SFDs for smaller objects ranges from
$-2.0$ to $-2.5$ (Grav et al. \cite{gra11}; Fraser et al. \cite{fra14}).  Of course, if the
planetesimal population evolves by collisions, new small objects are
likely produced as fragments of larger ones. This is guaranteed by the
above-mentioned shallow slopes of the planetesimal SFD, while it would
not hold true for a much steeper planetesimal SFD of the kind
suggested by Belton (\cite{bel15}).

\subsection{Number of collisions in the trans-planetary disk}

The number of disruptive collisions ($N_{\rm coll}$) for 67P can be
estimated using a particle-in-a-box formula (Wetherill \cite{wet67}),
\begin{equation}
N_{\rm coll} = \frac{N_p u T A_p}{V},
\end{equation}
where $N_p$ is the number of disk objects, $A_p = \pi (R_c+R_p)^2$ is
the cross section for a collision between 67P with average radius $R_c
= 2$\unit{km} and a partner with radius $R_p$, $u$ is the average
relative velocity, $T$ is the time interval, and $V$ is the volume of
the disk. Assuming viscous stirring of the disk by its largest members
(Levison et al. \cite{lev11}), we estimate $u = 0.6$\unit{km/s}, which is 10\%
of the typical orbital velocity in the disk. The volume is obtained by
considering a circular annulus with inner and outer radii of 15 and
30\unit{AU}, and a constant thickness of 4\unit{AU}.

As collision partners we consider disk objects with radius from
0.5\unit{km} up to the break point in the cumulative planetesimal SFD,
which occurs at $R_p \sim 50$\unit{km}. We take a power law with
constant index $\alpha$ for the cumulative planetesimal SFD and derive
the corresponding differential number $n(R_p) dR_p$ of disk objects in
a radius range of width $dR_p$. We replace $N_pA_p$ in Eq. (1) by the
integral
\begin{displaymath}
\int n(R_p) A_p dR_p \, ,
\end{displaymath}
taking the limits at 0.5\unit{km} and 50\unit{km}. The computation of
$n(R_p)$ will first be done as follows. At the end of the time
interval ($T = 0.4$\unit{Gy}) we use an estimate of $2\cdot10^{11}$ objects with $D >
2.3$\unit{km} in order to populate the scattered disk (Brasser and
Morbidelli \cite{bra13}). Populating the Oort cloud would seem to
require even more, but we prefer to be conservative. For the
cumulative SFD index $\alpha$ we consider two values: $-2.5$, which
would correspond to a relaxed, collisional asteroid population
(Dohnanyi \cite{doh69}), and $-2.0$, which is close to the value found
by a theoretical model (Johansen et al. \cite{joh15}) for small
planetesimals formed by pebble accretion. The latter would likely
characterize the initial state of the disk, and the former may
characterize the final state after collisional evolution.

For the sake of illustration, we use the above value for the
cumulative number with $D > 2.3$\unit{km} with the two $\alpha$ values
to characterize the disk throughout the time interval, even though it
explicitly refers to the final state. Our calculation then yields
$N_{\rm coll} \simeq 11$ and $9$ for $\alpha = -2.5$ and $-2.0$,
respectively. In that case, 67P-sized objects would undergo about ten
collisions on the average, and the chance to survive without any
collision, as estimated by a Poisson distribution, is less than
$10^{-4}$. Thus, for 67P to survive undisrupted, the initial $N_p$ 
would have to be at least a factor 10 smaller than we have
assumed. Hence, the final objects would have to be mostly collisional
fragments, and it is statistically unlikely for 67P to be among the
initial survivors.

Interestingly, the disk mass in the considered size range, assuming a
density of 1000~kg/m$^3$ (Brown \cite{bro13}), comes out as 6 or 18.5
Earth masses ($M_E$) for the two $\alpha$ values. We now consider
the contribution by objects with $D > 100$\unit{km}. The slope of the
cumulative SFDs for the trans-Neptunian populations and Trojan
asteroids for diameters larger than $\sim 100$\unit{km} ranges from
$-3.5$ to $-4.5$. We will use $\alpha = -4$ for the larger objects
(Gladman et al. \cite{gla01}; Fraser et al.  \cite{fra14}) while
keeping both limits discussed above for the smaller size range as in
the previous computation.  Very large objects (such as Pluto) are
insufficiently sampled by the adopted planetesimal SFD, but it is
expected that $\sim 1000$ Pluto-sized objects once existed in the
primordial disk (Levison et al. \cite{lev11}), thus contributing a few
Earth masses. With these assumptions, the total mass is compatible
with the inferred range from the Nice Model as stated above.

\begin{figure*}[h]
\includegraphics[width=9cm]{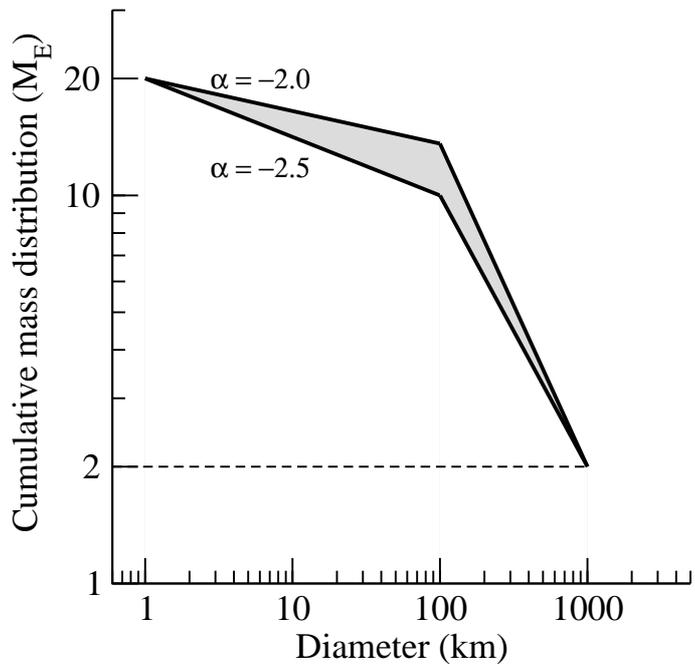}
\caption{A sketch of the primordial disk cumulative mass distribution
  (in units of Earth's masses), based on dynamical constraints and
  current observational constraints from Kuiper Belt objects and
  Trojan asteroids. We plot, as a function of diameter, the total disk
  mass contained in all larger objects. Two Earth masses have been
  allocated to objects larger than 1000$\unit{km}$. The total mass of
  objects larger than 1\unit{km} has been set to 20 Earth masses, at
  the low end of published primordial disk masses. The two solid
  curves correspond to slopes of $-2.0$ and $-2.5$ for the cumulative
  size-frequency distribution at $D < 100$\unit{km}.}
\label{fig_disk}
\end{figure*}

An alternative way of estimating $n(R_p)$ is as follows. We take
the minimum value of 20\unit{$M_E$} for the total disk mass. We
subtract 2\unit{$M_E$} for the Pluto-sized objects and consider
18\unit{$M_E$} to be partitioned between two populations with $D <
100$\unit{km} and $D > 100$\unit{km}. We distribute the 18\unit{$M_E$}
up to $D = 1000$\unit{km}, and as a result we find for $\alpha = -2.5$
that the number of objects with $D > 2.3$\unit{km} is $3\cdot10^{11}$,
yielding $N_{\rm coll} \simeq 18$, and for $\alpha = -2.0$ we find
$7\cdot10^{10}$, yielding $N_{\rm coll} \simeq 3$. In this case, even
for the shallower cumulative planetesimal SFD, the chance for survival
without any collision is only 5\%, and yet the planetesimal population
has an insufficient number of objects to feed the scattered disk.
Figure~\ref{fig_disk} illustrates the mass distribution used for the 
trans-planetary disk objects in this analysis.

\section{Escape from collisions}\label{sec:escape}

We conclude from the above that in both scenarios under consideration,
67P is likely to be collisionally evolved. To judge the chances
for 67P to be unaffected and primordial, we may consider a possible
gradient of collision probability across the disk. Its surface density
is usually modeled to vary as $r^{-1}$ with heliocentric distance
$r$. In addition, the thickness of the disk may have been proportional
to $r$, so that the number density may have varied as $r^{-2}$. This
would cause $N_{\rm coll}$ to be four times lower in the outermost
parts of the disk than in the innermost parts. Therefore, 67P may
possibly be a primordial survivor without being too exceptional, if it
formed in the outskirts of the disk, as possibly indicated by its high
D/H ratio (Altwegg et al. \cite{alt15}). In addition, an even lower
mass of the disk (Nesvorn\'y, pers. comm.), and a less dynamically
excited disk could also reduce the estimated collision rate by a few
times. In any case, it appears that the best chance for 67P to be
primordial is with a low-mass disk and a shallow size distribution of
the planetesimals (namely, $\alpha \lesssim -2$ for $D <
100$\unit{km}).

Our model features a common value of the encounter velocity for 
the trans-planetary disk objects. This is thought to be an average 
of the individual velocities, relevant to the average state of dynamical 
excitation of the disk. Of course, there are many encounters occurring 
at both lower and higher velocities, so we should consider the effect 
of this scatter. Since we used a cut-off of projectile diameters at 
$D = 1$\unit{km}, we cannot account for the fact that the size limit for 
disruptive collisions decreases with increasing velocity. However, this 
together with the fact that the collision frequency increases with velocity 
means that we have likely underestimated the number of disruptive 
collisions.

Another caveat concerns the timing of the giant planet
instability. Even though it is considered likely that the LHB was
triggered by the flux of asteroids and comets destabilized by the
planet migration, there remains a possibility that the instability
occurred earlier than assumed here. In such a case, the disk lifetime
may have been significantly less than 400\unit{My}, thus drastically
lowering the above estimates of $N_{\rm coll}$. However, a very early
instability would face other problems such as finding an alternate
explanation for the LHB.

We should also keep in mind that, whatever was the fate of the
planetesimals from which 67P stems, our estimates confirm the recent
conclusion by Morbidelli and Rickman (\cite{mor15}) that comets in
general did not escape collisional evolution in the early solar
system. This holds for all kinds of comets -- those coming from the
Oort cloud as well as those captured from the scattered disk or the
EKB.  Therefore, if 67P and other comets are found to retain
primordial structural features, it follows that these cannot have been
erased by the collisional history of the comets.

In this perspective, 67P appears to offer a precious test case, since
it has not stood out as very special among the JFCs. Thus, the study
of a collisional re-accretion origin for 67P is particularly
important.  For instance, should such an origin prove to be
inconsistent with the body of evidence from OSIRIS and other Rosetta
instruments in the sense that the primordial signatures (\eg,
layering and outgassing of super-volatiles) would have been erased, 
this would suggest that both 67P and
comets at large, are indeed primordial survivors. Such a conclusion 
would necessitate a reconsideration of the concepts around the 
early evolution of the solar system that were applied in this paper.

On the other hand, should it turn out that collisional re-accretion is
consistent with the observations of 67P including those features
thought to pre-date this accretion (\eg, layering and volatile abundances), the study of 67P
would still reveal important clues about early planetesimal formation,
independent of the role played by collisions in its later history.

\section{Discussion and conclusions}\label{sec:conclu}

We have shown that OSIRIS data helps to constrain the origin of comet
67P/Churyumov-Gerasimenko in several important ways. The bi-lobate
overall shape is best interpreted in terms of a contact binary, formed
by the collisional merger of two distinct bodies. Their properties
appear generally similar, but the smaller lobe may possibly show
evidence of damage due to an energetic impact, perhaps witnessed by
the observed large-scale fractures.

The large porosity derived from the bulk density of 470\unit{kg/m$^3$}
in fact requires accretional velocities for the two lobes
$\lta10-20$\unit{m/s}, and this leaves open two options for the origin
of 67P. Either it may be an undestroyed, primordial planetesimal that
accreted during the first $\sim10$\unit{My}, or it was formed by the
gentle re-accretion of fragments ejected from a larger parent body
upon a destructive collision at a later time (possibly, the last one
of a series of collisions). It is not possible yet to exclude either
of these scenarios. However, should the latter option be true, it must
have conserved the primordial signatures of 67P (both geomorphological
and chemical), and hence the comet remains a key witness of
planetesimal formation in the nascent solar system. We also emphasize
that the collisional scenario requires the bi-lobate nucleus to have
survived intact after it was formed, and thus the collision that led
to its formation must have occurred during the final stages of the
primordial disk or even during its dispersal.

The origin of internal layers remains elusive. However, it should be
noted that layering appears to be an inherent characteristic of the
two lobes prior to their accretion. According to Groussin et
al. (\cite{gro15}), the low compressional strength of 67P may
stimulate pressure-induced layering.  The early, gentle docking of two
km-sized planetesimals thus implies that their layering arose during
their primordial accretion.  Alternatively, the involvement of a
catastrophic disruption of a larger parent body (a few 10s\unit{km} at
most to avoid significant reduction of porosity by compaction) could
explain layering, if some internal evolution had taken place resulting
in a onion-shell structure, and the two lobes are large chunks of the
parent body rather than debris piles.

We also show that strong collisional evolution is generally to be
expected for the objects that occupied the early, trans-planetary
disk. While it is possible for 67P to have survived collisional
destruction in a safe niche under special assumptions about the disk,
we point out that 67P has not shown any evidence of being exceptional
among comets at large, and thus it should not have an exceptional
origin. However, this also means a challenge for the collisional
origin scenario. Low densities have been found for all Jupiter family
comets so far analyzed, and if these are generally produced by
destructive collisions involving large parent bodies, the latter must
have conserved deep surface layers in a primordial, porous state in
spite of $^{26}$Al-induced chemical differentiation (Prialnik et
al. \cite{pri04}).  Interestingly, several EKB objects in the size
range $D \sim 150-400$\unit{km} have been found to have a density of
$400-800$\unit{kg/m$^3$} (Brown \cite{bro13}), implying significant
porosity can be retained even on large objects.

Regarding the two main options, we conclude that the primordial
survivor option would likely call into question some of the current
concepts of solar system evolution, while the collisional fragment
option would not call for any such rethinking. The issue with the latter 
is to show it to be consistent with the observational evidence on 67P 
and other comets.

\begin{acknowledgements}

The authors wish to thank Hal Levison, Alessandro Morbidelli,
Katherine Kretke, David Nesvorn\'y, Alex Parker, Robert Grimm, Anders
Johansen for interesting discussions. We also thank the referee, David
O'Brien, for helpful suggestions. OSIRIS was built by a consortium led
by the Max-Planck-Institut f{\"u}r Sonnensystemforschung,
G{\"o}ttingen, Germany, in collaboration with CISAS, University of
Padova, Italy, the Laboratoire d’Astrophysique de Marseille, France,
the Instituto de Astrof{\'i}sica de Andalucia, CSIC, Granada, Spain,
the Scientific Support Office of the European Space Agency, Noordwijk,
Netherlands, the Instituto Nacional de T{\'e}cnica Aeroespacial,
Madrid, Spain, the Universidad Polit{\'e}chnica de Madrid, Spain, the
Department of Physics and Astronomy of Uppsala University, Sweden, and
the Institut f{\"u}r Datentechnik und Kommunikationsnetze der
Technischen Universit{\"a}t Braunschweig, Germany. The support of the
national funding agencies of Germany (DLR), France (CNES), Italy
(ASI), Spain (MEC), Sweden (SNSB), and the ESA Technical Directorate
is gratefully acknowledged. We thank the ESA teams at European Space
Astronomy Center, European Space Operations Center, and European Space
Research and Technology Center for their work in support of the
Rosetta mission. Special thanks are due to Ryszard Gabryszewski (PAS
Space Research Center, Warsaw) for permission to use the plot in
Figure~\ref{fig_rysz}. SM was supported by NASA Jet Propulsion
Laboratory Subcontract No. 1336850 to the Southwest Research
Institute. HR was supported by Grant No. 2011/01/B/ST9/05442 of the
Polish National Science Center.

\end{acknowledgements}

\bigskip
\bigskip

\noindent

\end{document}